\documentclass[%
 reprint,
 superscriptaddress,
 amsmath,amssymb,
 aps,longbibliography
]{revtex4-1}

\usepackage{graphicx}% Include figure files
\usepackage{dcolumn}% Align table columns on decimal point
\usepackage{bm}% bold math
\usepackage{hyperref}% add hypertext capabilities
\usepackage[mathlines]{lineno}% Enable numbering of text and display math
\usepackage[bottom]{footmisc}
\usepackage[super]{nth}%

\begin{document}
\raggedbottom

%\preprint{AIP/123-QED}
\title{Enhanced wave localization in multifractal scattering media}% Force line breaks with \\
% Yauyao
\author{Y. Chen}
\affiliation{Department of Electrical and Computer Engineering, Boston University, 8 Saint Mary's Street, Boston, Massachusetts 02215, USA}
% Sgrigno
\author{F. Sgrignuoli}
\affiliation{Institute of Applied Sciences and Intelligent Systems, National Research Council, Via Pietro Castellino, Naples, 80131, Italy}
% Sean
\author{Y. Zhu}
\affiliation{Division of Materials Science \& Engineering, Boston University, 15 Saint Mary's St. Brookline, Massachusetts, 02446, USA}
% T. Shubitidze
\author{T. Shubitidze}
\affiliation{Department of Electrical and Computer Engineering, Boston University, 8 Saint Mary's Street, Boston, Massachusetts 02215, USA}
% Luca
\author{L. Dal Negro}
\email{dalnegro@bu.edu}
\affiliation{Department of Electrical and Computer Engineering, Boston University, 8 Saint Mary's Street, Boston, Massachusetts 02215, USA}
\affiliation{Division of Materials Science \& Engineering, Boston University, 15 Saint Mary's St. Brookline, Massachusetts, 02446, USA}
\affiliation{Department of Physics, Boston University, 590 Commonwealth Avenue, Boston, Massachusetts 02215, USA}
%%%%%%%%%%%%%%%%%%%%%%%%%%%%%%%%%%%%%%%%%%%%%%%%%%%%%%%%%%%%%%%%%%%%%
%%%%%%%%%%%%%%%%%%%%%%%                  Abstract                      %%%%%%%%%%%%%%%%%%%%%%%%%%%%%
%%%%%%%%%%%%%%%%%%%%%%%%%%%%%%%%%%%%%%%%%%%%%%%%%%%%%%%%%%%%%%%%%%%%%
\begin{abstract}
In this paper we study the structural, scattering, and wave localization properties of multifractal arrays of electric point dipoles generated from multiplicative random fields with different degrees of multiscale correlations. Specifically, using the rigorous Green's matrix method, we investigate the scattering resonances and wave localization behavior of systems with $N=10^{4}$ dipoles and demonstrate an enhanced localization behavior in highly inhomogeneous multifractal structures compared to homogeneous fractals, or monofractals. We show distinctive spectral properties, such as the absence of level repulsion in the strong multiple scattering regime and power-law statistics of level spacings, which indicate a clear localization transition enhanced in non-homogeneous multifractals. Our findings unveil the importance of multifractal structural correlations in the multiple scattering regime of electric dipole arrays and provide an efficient model for the design of multiscale nanophotonic systems with enhanced light-matter coupling and localization phenomena beyond what is possible with traditional fractal systems.
\end{abstract}

\pacs{Valid PACS appear here}
\keywords{Suggested keywords}
%Use showkeys class option if keyword
\maketitle
%%%%%%%%%%%%%%%%%%%%%%%%%%%%%%%%%%%%%%%%%%%%%%%%%%%%%%%%%%%%%%%%%%%%%%%%%%%%
%%%%%%%%%%%%%%%%%%%%%%%%%%%%%%%%%%%%%%%%%%%%%%%%%%%%%%%%%%%%%%%%%%%%%%%%%%%%
% ==================================    Introduction    ==========================================================
%%%%%%%%%%%%%%%%%%%%%%%%%%%%%%%%%%%%%%%%%%%%%%%%%%%%%%%%%%%%%%%%%%%%%%%%%%%%
%%%%%%%%%%%%%%%%%%%%%%%%%%%%%%%%%%%%%%%%%%%%%%%%%%%%%%%%%%%%%%%%%%%%%%%%%%%%
\section{Introduction}
In recent years, self-similar structures attracted a significant interest in photonics and nano-optics technologies \cite{berry1979diffractals,soljacic2000self,bandres2016topological,aslan2016multispectral} adding novel functionalities to the manipulation of optical fields in complex media \cite{dal2012deterministic,torquato2018hyperuniform,segev2013anderson} beyond periodic \cite{joannopoulos1997photonic} or disordered systems \cite{wiersma2013disordered,lagendijk2009fifty}, with applications to resonant nano-devices and metamaterials \cite{shalaev2007optical,kabashin2009plasmonic,zheludev2012metamaterials}. 
Statistically-homogeneous optical media with fractal geometries, which naturally occur in a wide variety of physical systems including colloidal aggregates \cite{meakin1987fractal} and certain emulsions \cite{bibette1993structure}, motivated earlier studies on the single scattering properties of light in self-similar structures \cite{sorensen2001light}.      
Recently, the light transport through fractal structures beyond the single scattering regime has also been studied, leading to the demonstration of photon super-diffusive phenomena in novel L\'{e}vy glass optical media \cite{barthelemy2008levy,bertolotti2010engineering}. These are novel homogeneous materials containing TiO$_{2}$ nanoparticles and engineered fractal distributions of poly-dispersed glass spheres that enable control of photon scattering events in a self-similar environment with a small refractive index contrast.  
However, the complex geometry of many physical structures and multi-scale phenomena, ranging from fully developed turbulence and weather systems to the clustering of galaxies, network traffic, and the stock market, display very irregular fluctuations that cannot be adequately described by simple homogeneous fractal models. These more complex systems generally exhibit scaling properties that vary locally from point to point in space or time and require a continuum spectrum of fractal scaling exponents, i.e., a singularity spectrum, for their complete characterization \cite{frame2016fractal,Stanley,Mandelbrot2}.
These ideas let to the concept of multifractality \cite{stanley1988multifractal,chhabra1989direct,frisch1980fully} which, originally introduced in order to analyze multiscale energy dissipation in turbulent flows \cite{frisch1980fully,mandelbrot1974intermittent}, significantly broadened our understanding of complex structures in science and engineering \cite{Nakayama,jiang2019multifractal,riedi1999multifractal,albuquerque2003theory,sroor2019fractal,trevino2012geometrical}. Specifically, critical phenomena in disordered quantum and classical systems have been the subject of intense theoretical and experimental research leading to the discovery of multifractality in electronic and optical wave functions at the metal-insulator Anderson transition for conductors \cite{richardella2010visualizing,rodriguez2009multifractal,schreiber1991multifractal,evers2008anderson}, superconductors \cite{zhao2019disorder}, atomic matter waves \cite{chabe2008experimental}, and engineered nanophotonic structures  \cite{wang2018spectral,sgrignuoli2021optical,sgrignuoli2020multifractality,dal2021aperiodic,dalnegro2022waves}
Besides its fundamental interest, understanding the behavior of optical waves in strongly scattering multifractal media could offer a novel mechanism to localize and resonantly distribute classical and quantum light states at multiple length scales and to enhance light-matter coupling across broad frequency spectra. However, to the best of our knowledge, the distinctive multiple scattering and localization behavior of optical waves in multifractal arrays with controlled singularity spectra is still missing.

In this paper, we use the rigorous Green's matrix spectral method that enables a systematic investigation of complex scattering resonances and their spectral statistics in order to investigate the localization properties of light in open two-dimensional scattering arrays of electric dipoles ( i.e., systems with in-plane radiation losses) with fractal and multifractal geometrical arrangements. Specifically, by studying the Thouless conductance $g$ \cite{thouless1974electrons} and the ﬁrst-neighbor level spacing statistics for dfferent values of the optical density, our work demonstrates clear signatures of a broadband localization transition with significantly reduced $g$ values in multifractals compared to their monofractal counterparts. Moreover, we discover that multifractal arrays support a significantly larger density of eigenmodes in the localization regime. Finally, we address the spectral statistics of level spacings and show a crossover from a level repulsion behavior described by the Gaussian unitary ensemble (GUE) of random matrices \cite{Mehta} at low optical density to a level clustering behavior with power-law level spacing distributions at large optical density. Our findings shows the enhanced localization properties of highly-inhomogeneous multifractal arrays compared to traditional fractal systems and provide yet-unexplored possibilities to systematically exploit multifractality as a novel strategy for the engineering of novel nanophotonic systems with broadband localization properties for optical sensing, random lasing, and multi-spectral devices.  

Our paper is organized as follows. In section \ref{fractals}, we introduce background concepts on fractals and multifractals. In section \ref{arrays}, we discuss the generation of the investigated fractal and multifractal arrays and introduce their structural and wave diffraction properties. In section \ref{spectral}, we present our results on the spectral and localization properties of the scattering resonances of fractals and multifractals and we draw our conclusions in section \ref{conclusions}. Finally, in Appendix \ref{MFA} we review the relevant concepts of multifractal analysis and in Appendix \ref{diffraction} the single scattering (i.e., diffraction) regime of fractals and multifractals.  

\section{Fractals and multifractals}\label{fractals}
Fractal objects are characterized by non-integer fractal dimensions and exhibit power-law scaling of structural (i.e., density-density correlation, structure factor) and dynamical (i.e., density of modes, spectral functions) properties \cite{gouyet1997physics}. In fractal systems, the exponent of the power-law scaling of the mass with the system size does not coincide with the Euclidean dimension. This property implies very large local density ﬂuctuations and high lacunarity \cite{gouyet1997physics}, leading to the existence of both very dense and very empty regions. The fractal dimensions of physical objects is operationally defined using the box-counting method \cite{falconer2004fractal}. In this approach, the space embedding the fractal is sub-divided into a hyper-cubic grid of boxes (i.e., cells) of linear size $\epsilon$ (i.e., line segments in the case of one-dimensional objects, squares in two dimensions, cubes in three dimensions, and so on). For a given box of size $\epsilon$, the minimum number of boxes $N(\epsilon)$ needed to cover all the points of the object is determined and this procedure is repeated for several box sizes. Finally, the (box-counting) fractal dimension $D_{0}$ is determined from the power-law scaling:
\begin{equation}
N(\epsilon)\propto{\epsilon^{-D_{0}}}
\end{equation}

The procedure can be extended to any suitable measure defined on a set, and the fractal properties of the measure (e.g., the number of components or the mass density of an object) are obtained by studying the scaling behavior of its moments with respect to the size of covering partitions of the set. Alternatively, for arrays of point particles we can determine the fractal dimension by drawing a sphere of radius $r$ and computing the total number (or the mass)  of the particles included in this sphere denoted by $N(r)$. 
Moreover, if randomness is involved in the fractal object, we consider the scaling of $N(r)$ over spheres with different centers. Then the (averaged) fractal dimension $D_{m}$ is determined from the scaling law:
\begin{equation}
\langle{N(r)}\rangle\,\propto\,{r^{D_{m}}}
\end{equation}
where $\langle\ldots\rangle$ denotes the average over different spheres with radius $r$ and the fractal dimension $D_{m}$ is also called the mass fractal dimension \cite{gouyet1997physics}. 
The box-counting or mass fractal dimensions introduced above provide a concise description of how the size or mass of a fractal vary with respect to the magnification scale $\epsilon$. Objects that are uniquely described by a constant (scale-independent) fractal dimension are called homogeneous fractals, or monofractals, and possess features that repeat identically at every scale i.e., they exhibit scale-invariance symmetry or self-similarity over a large range of scales. 
The relevance of fractals to physical sciences and other disciplines (i.e., economics) was originally pointed out by the pioneering work of Mandelbrot \cite{mandelbrot1982fractal}. More recently, the concept of multifractals, or inhomogeneous fractals, has been introduced to characterize complex systems with space or time dependent self-similar properties and a rigorous multifractal formalism has been developed to quantitatively describe their local scaling \cite{mandelbrot1974intermittent,frisch1980fully,stanley1988multifractal}, which is briefly reviewed in Appendix \ref{MFA}. 

Multifractal behavior is characteristic of random multiplicative processes, such as the ones arising when multiplying together strings of random numbers. 
Examples of multifractal structures and phenomena are commonly encountered in dynamical systems theory (e.g., strange attractors of nonlinear maps), physics (e.g., diffusion-limited aggregates, fluid dynamics), engineering (e.g., random resistive networks, image analysis), geophysics (e.g., rock shapes, creeks), atmospheric science (analysis of rain and clouds), as well as in statistics and finance (e.g., extreme value theory, stock markets fluctuations) \cite{harte2001multifractals,falconer2004fractal,stanley1988multifractal,riedi1999multifractal,frisch1980fully,jiang2019multifractal}. 
Generally, when dealing with multifractals on which a local measure $\mu$ is defined (i.e., a mass density, a velocity, an electrical signal, or some other scalar physical parameter defined on the fractal object), the (local) singularity strength $\alpha(x)$ of the multifractal measure $\mu$ obeys the more complex scaling law: 
\begin{equation}
\mu[B_{x}(\epsilon)]\propto{\epsilon^{\alpha(x)}}
\end{equation}
where $B_{x}(\epsilon)$ denotes a ball (i.e., an open interval) of small radius $\epsilon$ centered at $x$. 
The exponent $\alpha(x)$ measures the local singularity of the measure, i.e., the smaller the exponent $\alpha(x)$, the more singular is the measure around $x$. The multifractal spectrum $f(\alpha)$, also known as the singularity spectrum, characterizes the statistical distribution of the singularity exponent $\alpha(x)$ of the multifractal measure \cite{falconer2004fractal,harte2001multifractals}. Moreover, if we cover the support of the measure $\mu$ with balls of size $\epsilon$, the number of balls $N_{\alpha}(\epsilon)$ which, for a given $\alpha$, scales like $\epsilon^{\alpha}$, is given by:
\begin{equation}
N_{\alpha}(\epsilon)\propto{\epsilon^{-f(\alpha)}}
\end{equation}
It has been established that in the limit of vanishingly small $\epsilon$, $f(\alpha)$ coincides with the fractal dimension of the set of 
all points $x$ with the same scaling index $\alpha$. The spectrum $f(\alpha)$ was originally introduced by Frisch and Parisi \cite{frisch1980fully} to investigate the energy dissipation of turbulent fluids. From a physical point of view, the multifractal spectrum is a quantitative measure of structural inhomogeneity and it is well suited for characterizing complex spatial signals because it can efficiently resolve their local fluctuations. In the case of multifractal measures with a recursive multiplicative structure, such as the ones investigated in this work, the multifractal spectrum can be calculated analytically \cite{martinez1990clustering}. However, in general it is obtained numerically by implementing the formalism reviewed in Appendix \ref{MFA} using, for example, the efficient and accurate method developed by Chhabra and Jensen \cite{chhabra1989direct}.

\section{Multifractal arrays} \label{arrays}
The multifractal scattering arrays investigated in this paper are generated by considering random processes based on hierarchical probability cascades. A multiplicative cascade model \cite{anitas2019small}  is constructed by first dividing a square into four equal squares. To each of the sub-squares, one assigns the probabilities $p_{i}\in[0, 1]$ with $i = 1, 2, 3, 4$. This constitutes the first iteration of the process ($n=1$). At the second iteration ($n = 2$), each of the four sub-squares is further divided in four squares, and 
the probabilities associated with each sub-division are multiplied in random order with the ones of the previous iterations.
At the third iteration, one performs a similar division into sub-squares and to each of them assigns the probabilities in random permutations from the previous iterations, i.e., $n = 1$ and $n = 2$. The resulting  multiplicative cascade multifractal distribution is the probability field obtained in the limit of a large number of iterations \cite{martinez1990clustering}.
The probability value attached to a square region is the product of the $p_{i}$'s of the square and all its ancestors at previous generations and the distribution of cell values strictly depends on the initial choice of the probability vector $p_{i}$. 
Because the random numbers in the probability fields are generated multiplicatively, the corresponding process is a multiplicative random process that is generally non-Gaussian. 
Random point patterns (i.e., point processes) with multifractal scaling properties induced by the  probability fields introduced above are generated by distributing $N$ particles on the underlying square lattice with probabilities that are simply proportional to the square-cell values. We used a Monte Carlo rejection scheme for generating multifractal arrays with $N=10^{4}$ point dipoles \cite{martinez1990clustering}.

\begin{figure*}[t!]
	\centering
	\includegraphics[width=\linewidth]{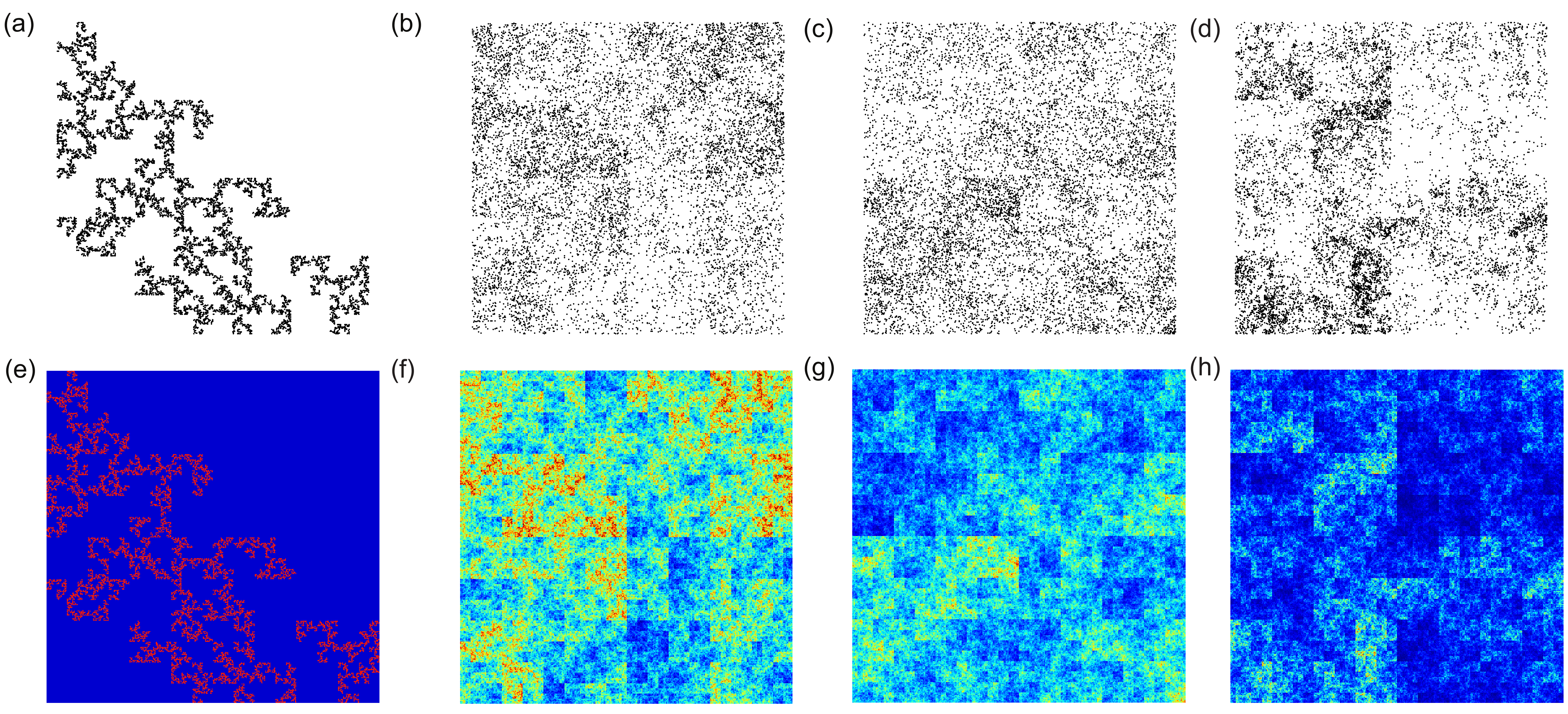}
    \caption{(a-d) Point patterns and (e-h) probability matrices for single realization multifractal structures. Panels (a,e), (b,f), (c,g), and (d,h) refer to the single realization of multifractal patterns composed by 10000 elements and generated with probabilities $p=[1,1,1,0]$, $p=[1,1,0.75,0.5]$, $p=[1,0.75,0.75,0.5]$, $p=[1,0.75,0.5,0.25]$, respectively.}
	\label{Fig1}
\end{figure*}

\begin{figure}[t!]
	\centering
	\includegraphics[width=\linewidth]{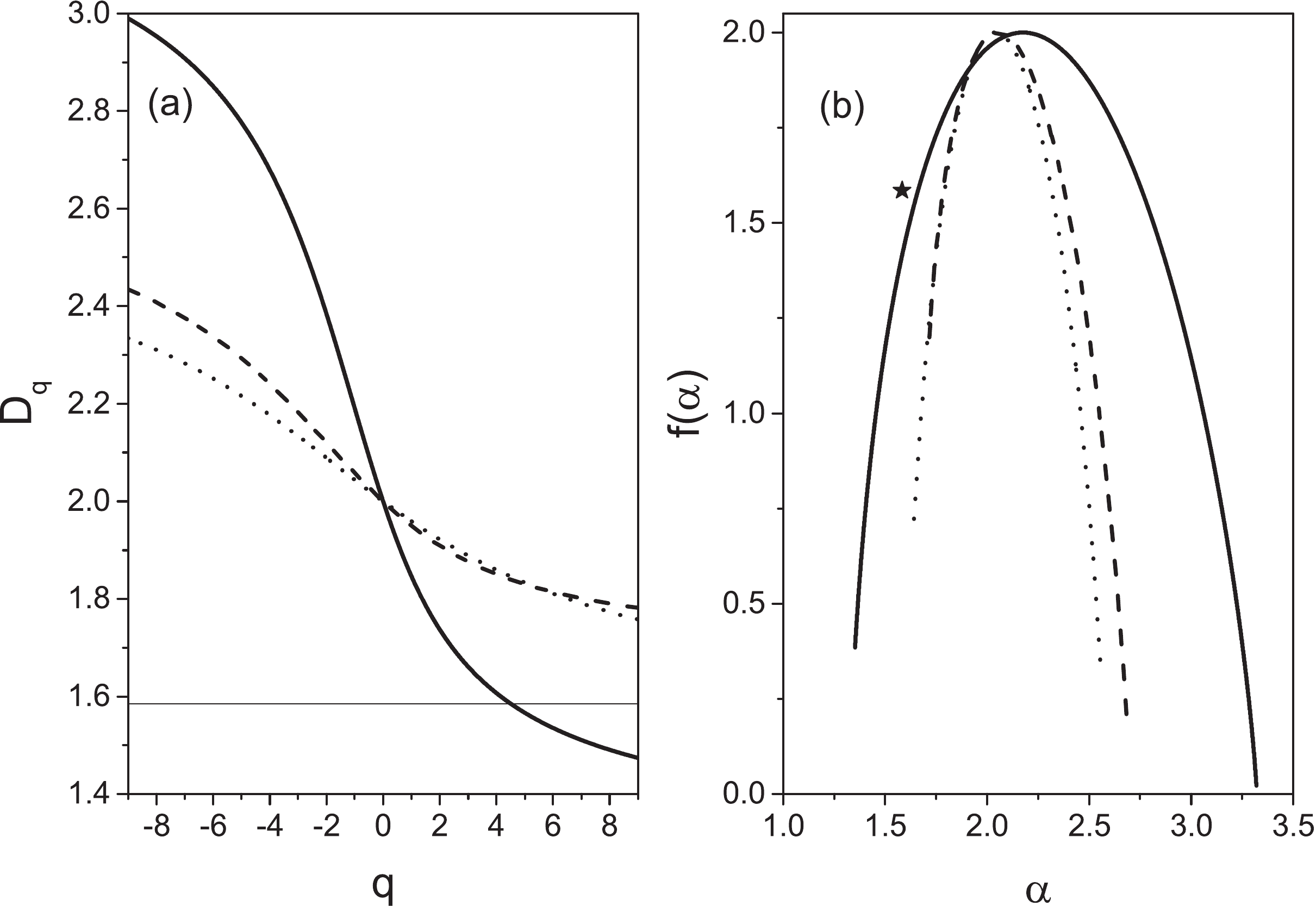}
    \caption{(a) Spectra of generalized dimensions $D_q$ and (b) multifractal singularity spectra $f(\alpha)$ for multifractal structures generated with probability  vectors $p=[1,1,1,0]$ (thin solid line and star symbol), $p=[1,1,0.75,0.5]$ (dotted line), $p=[1,0.75,0.75,0.5]$ (dashed line), and $p=[1,0.75,0.5,0.25]$ (thick solid line).}
	\label{Fig2}
\end{figure}

In Fig. \ref{Fig1} we show the investigated arrays in panels (a-f) and the corresponding probability fields in panels (c-h). The arrays are constructed from the four probability vectors reported in the caption. These are chosen so that the array in panel (a) is a statistically homogeneous fractal while the arrays in panels (b-f) are multifractals with an increasing degree of spatial non-uniformity.
The multifractal spectra of particle arrays generated from the multiplicative cascade model are computed analytically in the limit of systems with large size $L$.
In particular, the spectrum of generalized dimensions has the following analytical expression \cite{martinez1990clustering}:
\begin{equation}
D_{q}=\frac{1}{1-q}\log_{2}(f_{1}^{q}+f_{2}^{q}+f_{3}^{q}+f_{4}^{q})
\end{equation}
where $f_{i}=p_{i}/\sum_{i=1}^{4}p_{i}$ with $i=1,2,3,4$.
The corresponding multifractal spectrum $f(\alpha)$ is then calculated according to Eq. \ref{impo1} where $\tau(q)=(1-q)D_{q}$. 
The spectra of generalized dimensions $D_{q}$ and the multifractal spectra $f(\alpha)$ for the investigated arrays shown in Fig. \ref{Fig1} are displayed in Fig. \ref{Fig2}. The results show clearly a transition from an homogeneous fractal structure, featuring a constant box-counting $D_{0}\approx{1.58}$ and $f(\alpha)$ supported only by a single point, to more inhomogneous multifractals characterized by increasingly broader $f(\alpha)$ spectra for decreasing amplitudes of the initial probability vectors.

\begin{figure}[t!]
	\centering
	\includegraphics[width=\linewidth]{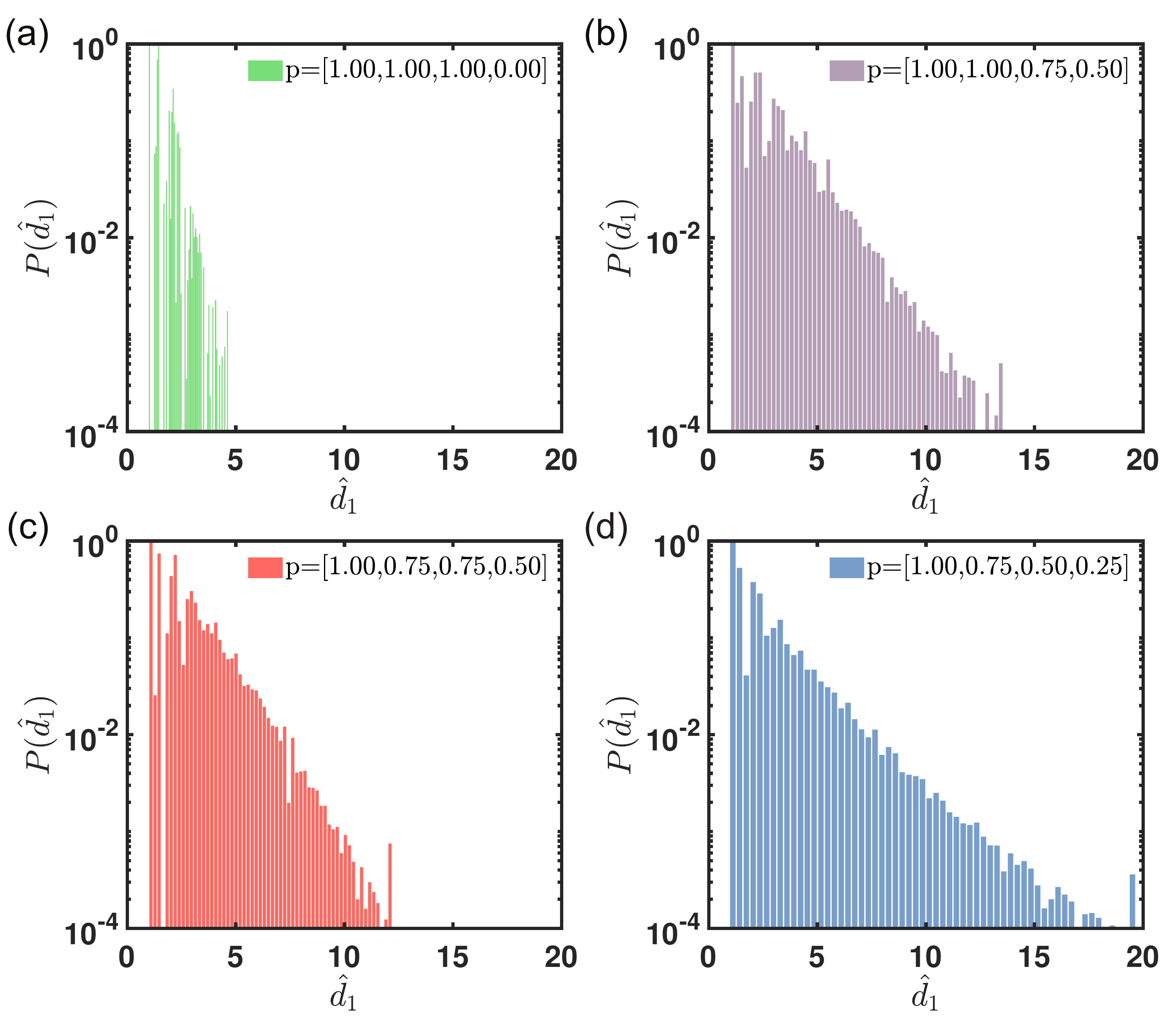}
    \caption{Normalized histograms of first-neighbor distances for multifractal structures generated with probabilities (a) $p=[1,1,1,0]$, $p=[1,1,0.75,0.5]$, $p=[1,0.75,0.75,0.5]$, $p=[1,0.75,0.5,0.25]$.}
	\label{Fig3}
\end{figure}

In Fig. \ref{Fig3} we further characterize the geometrical structure of the investigated arrays by analyzing the normalized probability distributions of first-neighbor distances of the particles averaged over 25 realizations of the disorder. Our findings show that multifractal arrays develop significantly broader probability distributions compared to the monofractal case shown in panel (a), consistently with the increased degree of spatial non-uniformity associated to their broad multifractal spectra. Specifically, we found that the arrays with the broader support of the spectrum $f(\alpha)$ also feature the larger range of distances in the distribution histograms shown in Fig. \ref{Fig3}.

In Fig. \ref{Fig4} we address the single scattering wave properties of the analyzed fractal and multifractal structures by showing in panel (a) the azimuthally-averaged structure factors and in panel (b) the corresponding pair correlation functions on double logarithmic scales. It is well-known that in fractals and multifractals these quantities display 
a power-law decay with a slope determined by the average fractal dimension \cite{freltoft1986power,dimon1986structure} (see Appendix \ref{diffraction}). In particular, for a homogeneous fractal, the pair-correlation scales as $g_{2}(r)\propto{r^{-\beta}}$ and similarly for the structure factor $S(q) \propto {|q|^{-(d-\beta)}}$ where $0<\beta<d$, consistently with the data presented in Fig. \ref{Fig4}. In particular, for the monofractal structure (green lines) the slopes extracted from the linear fit of the structure factor and the $g_{2}(r)$ in double logarithmic scale yield exactly the box-counting fractal dimension $D_{0}=1.58$ with $\beta=d-D_{0}$. However, the exact relation between the exponent $\beta$ and the fractal dimension $D_{0}$ is non-trivial for more general fractals and multifractals as it depends on the process of structure formation as well as their spatial distributions \cite{soneira1977there}. 
In all cases, it is important to realize that realistic finite-size systems exhibit lower and upper bound cutoffs in their fractal or multifractal nature. This implies that the power-law decays are observed only over a small range of scales and the correlation $g_{2}(r)$ eventually converges to $1$ at large $r$, as shown in Fig. \ref{Fig4}. We provide additional information on the single scattering properties of fractals and multifractals in Appendix \ref{diffraction} and focus next on the multiple scattering regime.

\section{Spectral and localization properties of multifractal arrays} \label{spectral}
% 2D Hankel Green
We now investigate the wave transport and localization properties of TM-polarized electric dipoles that are spatially arranged as in Fig.\ref{Fig1}. Multiple scattering effects in two spatial dimensions (i.e., for cylindrical waves) are studied by analyzing the spectral properties of the Green\textsc{\char13}s matrix defined as:
%%%%%%%%%%%%%%%%%%%%%% Green Matrix %%%%%%%%%%%%%%%%%%%%%% 
\begin{equation}\label{Green}
G_{ij}=i\left(\delta_{ij}+\tilde{G}_{ij}\right)
\end{equation}
%%%%%%%%%%%%%%%%%%%%%%%%%%%%%%%%%%%%%%%%%%%%%%%%%%%
where the elements $\tilde{G}_{ij}$ are given by \cite{RusekPRE2D}:
%%%%%%%%%%%%%%%%%%%%%% Green Our %%%%%%%%%%%%%%%%%%%%%%%
\begin{eqnarray}\label{GreenOur}
\begin{aligned}
\tilde{G}_{ij}=\frac{2}{i\pi}K_0(-ik_0|\textbf{r}_i-\textbf{r}_j|)
\end{aligned}
\end{eqnarray}
%%%%%%%%%%%%%%%%%%%%%%%%%%%%%%%%%%%%%%%%%%%% %%%%%%
and $K_0(-ik_0|\textbf{r}_i-\textbf{r}_j|)$ denotes the modified Bessel function of the second kind, $k_0$ is the wavevector of light, and $\textbf{r}_i$ specifies the position of the $i$-th scattering dipole in the array. The non-Hermitian matrix (\ref{Green}) describes the electromagnetic coupling among the scatterers and the real and imaginary part of its complex eigenvalues $\Lambda_n$ ($n\in$ 1, 2, $\cdots$N) correspond to the detuned frequency $(\omega_0-\omega)$ and  decay rate $\Gamma_n$ (both normalized to the resonant width $\Gamma_0$ of an isolated dipole) of the scattering resonances of the system \cite{RusekPRE2D,Lagendijk}. This formalism accounts for all the multiple scattering orders and enables the systematic study of the scattering properties of 2D waves with an electric field parallel to the invariance axis of the scatterers \cite{Leseur}. Even though the 2D model in (\ref{Green}) does not take into account the vector nature of light \cite{fab2019localization,SkipetrovPRL}, it still provides useful information on the localization properties of light in 2D disorder media \cite{RusekPRE2D} and aperiodic structures \cite{SgrignuoliPinwheel}, transparency in high-density hyperuniform materials \cite{Leseur}. Moreover, it correctly describes the coupling between one or several quantum emitters embedded in structured dielectric environments  \cite{Caze,Bouchet}. 

To investigate the spectral properties of the designed arrays, we analyze the distributions of the complex eigenvalues and representative scattering resonances, the behavior of the Thouless number $g$ as a function of the frequency $\omega$ and study its minimum value for each considered optical density $\rho\lambda^2$, and the level spacing statistics for different values of $\rho\lambda^2$. Here, $\rho$ denotes the number of scatterers per the unit area, and $\lambda$ is the optical wavelength. These spectral information are derived by numerically diagonalizing the $N\times N$ Green\textsc{\char13}s matrix (\ref{GreenOur}). 

% Discussion figure 4: eigenvalue distributions
At low optical density (i.e., $\rho\lambda^2=10^{-6}$) all the investigated systems are in the diffusive regime. Accordingly, their eigenvalue distributions, color-coded according to the $\log_{10}$ of the modal spatial extent (MSE), do not show the formation of any long-lived scattering resonances, as shown in Fig.\,(\ref{Fig5})\,(a-b). The MSE parameter quantifies the spatial extension of a given scattering resonances $\Psi_i$ of the system and it is defined as \cite{SgrignuoliACS}:
\begin{equation}
\text{MSE}=\left(\displaystyle\sum\limits_{i=1}^{N} \left|\Psi_i\right|^2\right)^2\Big/\displaystyle\sum\limits_{i=1}^{N}  \left|\Psi_i\right|^4
\end{equation}
On the other hand, at large optical density $\rho\lambda^2=50$, we observe the appearance of long-lived scattering resonances forming near $\omega\approx-2$, as visible in Fig.\,\ref{Fig5}\,(e-h). Consistently, the corresponding Green\textsc{\char13}s matrix eigenvectors, reported in Fig.\ref{Fig6}, are spatially localized over small clusters of dipoles demonstrating the formation of Efimov-type few-body scattering resonances occurring due to locally symmetric particle clusters distributed across the investigated structures \cite{fab2019localization, sgrignuoli2022subdiffusive}. Moreover, Fig.\,\ref{Fig5} also shows  the formation of critical scattering resonances for point patterns with multifractal properties. Critical modes are spatially extended and long-lived resonances with spatial fluctuations at multiple length scales characterized by a power-law scaling behavior \cite{macia1999,RyuPRB}, which are the results of the effect of local correlations on wave interference across the structures \cite{sgrignuoli2020multifractality}.

The Thouless number $g$ as a function of $\omega$ is evaluated as:
\begin{equation}\label{Thouless}
g(\omega) =\frac{\overline{\delta\omega}}{\overline{\Delta\omega}}=\frac{(\overline{1/\Im[\Lambda_n]})^{-1}}{\overline{\Re[\Lambda_n]-\Re[\Lambda_{n-1}]}}
\end{equation}
following the same procedure as in our previous work \cite{fab2019localization,Sgrignuoli_PRB2020,sgrignuoli2022subdiffusive,SgrignuoliPinwheel}. In particular, we have sampled the real parts of the eigenvalues of the Green\textsc{\char13}s matrix in 5000 equi-spaced intervals and we computed eq.\,(\ref{Thouless}) in each frequency sub-interval. The symbol $\overline{\{\cdots\}}$ in eq.\,(\ref{Thouless}) denotes the sub-interval averaging operation, while $\omega$ indicates the central frequency of each sub-interval. We have verified that the utilized frequency sampling resolution does not affect the presented results. 

Figure\,\ref{Fig7} shows the results of the Thouless number analysis in both the dilute and multiple scattering regimes. Consistently with the low value of the optical density, we found that the Thouless number is always larger than the one, indicating a diffusive regime \cite{SgrignuoliPinwheel,SkipetrovPRL}. At larger optical density, the Thouless number shows a completely different behavior with clear spectral ranges where  $g$ drops below one, indicating the onset of light localization. We remark that the long-lived scattering resonances that are spatially confined over few scatterers appear at the frequency positions where the Thouless number becomes lower than one. Interestingly, we found that point patterns with multifractal geometrical feature enhanced localization effects characterized by significantly smaller Thouless numbers compared to the investigated monofractal structures generated with a probability $p$ equal to $[1 ,1, 1, 0]$ (i.e., pastel green markers). This outcome is more evident by looking at the insets in Fig.\ref{Fig7}\,(e-h), showing a zoom-in view of the Thouless number in the frequency range $\omega\in[-10, 2]$.

To obtain additional insights on the localization behavior of multifractal structures, we analyze the minimum value of the Thouless number as a function of  $\rho\lambda^2$. Specifically, we have evaluated $g=g(\omega)$ by using eq.\,(\ref{Thouless}) for each $\rho\lambda^2$ value and we have repeated this procedure for 25 different point pattern realizations for each investigated structures. The circle markers and the error bars in Fig.\,\ref{Fig8}\,(a) are the averaged values and the standard deviations corresponding to this ensemble-averaged operation, in the follow identifies by the symbol $\langle\cdots\rangle_e$. Specifically, all these curves cross the delocalization-localization threshold value $g =1$ at the same optical density range, i.e., $\rho\lambda^2\in[10^{-1}, 1]$, demonstrating the diffusion to localization transition. In order to understand the observed transition in these novel structures we must consider the ratio $\xi/{L}$, where $L$ is the system size and $\xi$ identifies the localization length provided by \cite{Sheng}: 
\begin{equation}\label{xi}
\xi\sim l_t\exp[\pi \Re(k_e)l_t/2]
\end{equation}
with $l_t$ the transport mean free path and $\Re(k_e)$ the real part of the effective wavenumber in the medium. Although the numerical factor in eq.\,(\ref{xi}) may not be accurate \cite{Sheng,Gupta}, it nevertheless tells us that the localization length in 2D systems is an exponential function of $l_t$ and can be extremely large in the weak scattering regime (i.e., at low optical density). Consistently, we found that $\xi/L$ is very large (i.e., $\xi/L\gg1$) at low optical density indicating diffusion, while it becomes smaller than one, signaling localization, at the larger values of $\rho\lambda^2$ used in our analysis. Moreover, the investigated systems feature broadband spectra of localized optical resonances with distinctive scaling properties associated to their geometrical multifractality. This result is shown in Fig.\,\ref{Fig8}\,(b) where the width of the frequency range $\Delta\omega_{loc}$ corresponding to a Thouless number lower than 1 is reported for different optical densities. Moreover, Fig.\,\ref{Fig8}\,(c) also displays the ensemble-averaged number of localized scattering resonances $\langle N_{loc} \rangle_e$, demonstrating broadband enhancement of light-matter interactions produced by engineered geometrical supports with multifractal features. 

In order to better understand the nature of localization in the investigated systems we perform a statistical analysis of the level spacing that is often utilized to identify different transport
regimes in closed (Hermitian) and open systems \cite{Skipetrov2015,sgrignuoli2020multifractality,fab2019localization,sgrignuoli2022subdiffusive}. In closed random
systems, established results from random matrix theory
(RMT) predict the suppression of level repulsion in the
presence of localized states \cite{Mehta}. In these systems, spatially
separate, exponentially localized modes hardly influence
each other and can coexist at energies that infinitely
close. 
The transition from diffusion to localization is confirmed by the switching from level repulsion to level clustering of the quantity $\langle P(\hat{s})\rangle_e$ as a function of $\rho\lambda^2$, which is demonstrated in Fig.\,\ref{Fig9}. Here, $P(\hat{s})$ denotes the probability density function of the first-neighbor level spacing distribution of the complex eigenvalues of the Green\textsc{\char13}s matrix \cite{Escalante}. It is well-established that the suppression of the level repulsion (i.e., $P(\hat{s})\rightarrow$0 when $\hat{s}$ goes to zero) indicates the transition into the localization regime for both scalar and vector waves in two-dimensional and three-dimensional disordered systems \cite{Skipetrov2015,Escalante,Mondal} as well as non-uniform aperiodic deterministic structures \cite{fab2019localization,wang2018spectral,Sgrignuoli_PRB2020}. When $\rho/k_0^2 = 10^{-6}$, the distribution of the level spacing of the investigated arrays show an excellent agreement with the  Gaussian unitary ensemble (GUE) formula \cite{Mehta,Haake}:
\begin{equation}\label{GUE}
P(\hat{s})=\frac{32~\hat{s}^2}{\pi^2}~e^{-4\hat{s}^2/\pi}
\end{equation}
We emphasize that the black-dashed lines in Fig.\,\ref{Fig9}\,(a-d) are not the results of a numerical fitting procedure but are simply obtained by using eq.\,(\ref{GUE}). This distribution falls off quadratically for $\hat{s}\rightarrow0$ \cite{Mehta,Haake}, demonstrating that the eigenvalues of the investigated structures exhibit quadratic level repulsion in the low scattering density regime. Interestingly, the GUE distribution (\ref{GUE}) has also been discovered in the spacing of the non-trivial zeros of the Riemann's zeta function \cite{Odlyzko}, whose properties are intimately related to the distribution of prime numbers. Approaching the threshold of the discovered transition, the quantity $\langle P(\hat{s})\rangle_e$ manifests level repulsion described by the \textit{critical} cumulative probability defined as \cite{Zharekeshev}:
\begin{equation}\label{CriticalDistribution}
I(s)=\exp\left[\mu-\sqrt{\mu^2+(A_cs)^2}\right]
\end{equation} 
where $\mu$ and $A_c$ are fitting parameters, as shown in Fig.\,\ref{Fig9}\,(e-h). The critical cumulative probability was successfully applied to describe the energy level spacing distribution of an Anderson Hamiltonian containing $10^6$ lattice sites at the critical disorder value, i.e., at the metal-insulator threshold where it is known that all the wave functions exhibit multifractal scaling properties \cite{Zharekeshev}. Our findings demonstrate the applicability of critical statistics to this new class of multifractal structures reflecting the formation of critically localized eigenmodes with self-similar scaling properties. 

In contrast, the level spacing distributions of all the investigated structures are well-reproduced by the inverse power law scaling curve $P(s)\sim s^{-\beta}$ (see black-dotted lines in Fig.\,\ref{Fig9}\,(i-l)) in the strong multiple scattering regime ($\rho/k_0^2 = 50$). In the context of random matrix theory, it has been demonstrated that the power-law distribution describes complex systems with multifractal spectra that produce uncountable sets of hierarchical level clustering \cite{Cvitanovic,Geisel}. Moreover, this power-law scaling is related to anomalous diffusion. Anomalous diffusion is a transport regime in which the width of a wavepacket $\sigma^2$ increases upon propagation according to $t^{2\nu}$ with $\nu\in[0,1]$ \cite{Cvitanovic}. Such a behavior was observed in different one and two dimensional aperiodic systems \cite{Geisel,Guarneri,Benza}, and, more recently, in three-dimensional scattering arrays designed from sub-random sequences \cite{Sgrignuoli_PRB2020} and stealthy hyperuniform disordered systems \cite{sgrignuoli2022subdiffusive}. The anomalous exponent $\nu$ is related to the parameter $\beta$ through the relation $\nu=(\beta-1)/d$ where $d$ is the Euclidean dimension of the system \cite{Cvitanovic,Geisel}. By substituting the values of $\beta$ obtained from the numerical fitting of the data in Figure\,\ref{Fig9}\,(i-l), we find that the exponent $\nu$  is equal to $0.11\pm0.03$, $0.08\pm0.01$, $0.08\pm0.01$, and $0.06\pm0.01$ for structures generated with the probabilities vector $p$ equal to [1,1,1,0], [1,1,0.75,0.5], [1,0.75,0.75,0.5], and [1,0.75,0.5,0.25], respectively. These values are always much smaller than 0.5, consistently with the critical localization regime achieved at the highest optical density that we have investigated. 

\section{Conclusions}\label{conclusions}
%%%%%%%%%%%%%%%%%%		  
In conclusion, we have addressed the structural, spectral and localization properties of multifractal arrays of electric point dipoles with different degrees of multiscale structural correlations. We systematically studied the multiple scattering properties of classical waves using the spectral Green's matrix method and demonstrated an enhanced localization behavior in non-homogeneous multifractal structures compared to homogeneous fractals. In particular, we found strongly reduced Thouless numbers in the localization regime of multifractals accompanied by a larger number of localized scattering resonances with critical level spacing statistics at large optical density.
Our results demonstrate the importance of tailoring multifractal correlations in the multiple scattering regime for the engineering of novel nanophotonic systems with broadband localization properties for optical sensing, lasing, and multi-spectral devices.  

%%%%%%%%%%%%%%%%%%%%%%%%%%%%%%%%%%%%%%%%%%%%%%%%%%%%%%%%%%
\begin{acknowledgments}
L.D.N. acknowledges the support from the National Science Foundation (ECCS-2110204).
\end{acknowledgments}
%%%%%%%%%%%%%%%%%%%%%%%%%%%%%%%%%%%%%%%%%%%%%%%%%%%%%%%%%%%%%%%%%%%%%%%%%%%%
%%%%%%%%%%%%%%%%%%%%%%%%%%%%%%%%%%%%%%%%%%%%%%%%%%%%%%%%%%%%%%%%%%%%%%%%%%%%
% ==================================    Methods Greens    ==========================================================
%%%%%%%%%%%%%%%%%%%%%%%%%%%%%%%%%%%%%%%%%%%%%%%%%%%%%%%%%%%%%%%%%%%%%%%%%%%%
%%%%%%%%%%%%%%%%%%%%%%%%%%%%%%%%%%%%%%%%%%%%%%%%%%%%%%%%%%%%%%%%%%%%%%%%%%%%

\appendix

\section{Multifractal analysis}\label{MFA}
Multifractal analysis is based on the scaling properties of the partition function $Z_{q}(\epsilon)$. We consider an object on which a measure $\mu$ is distributed with constant density so that its multifractal properties are manifested purely in the scaling of its geometry. In this case the measure $\mu$ can be regarded simply as the mass density of the fractal object and the multifractal spectrum describes the geometrical support itself.
The widespread box-counting method considers a uniform square grid of boxes with linear size $\epsilon$ and then introduces the local measure $\mu_{i}$ as the proportion of total mass of the object inside the $i$-th box of size $\epsilon$. Then the partition function is defined as:
\begin{equation}\label{partition}
Z_{q}(\epsilon)=\sum_{i=1}^{N}\mu_{i}^{q}(\epsilon)
\end{equation}
The expression above denotes the sum of the $q$-th moments of the local measures over all the boxes needed to cover the support and is also known as the moment sum. We note that the higher the values of $q$ in \ref{partition}, the more dense are the selected regions. 

Multifractal analysis assumes a power-law behavior for the partition function in the limit $\epsilon\rightarrow{0}$ (or $N\rightarrow\infty$) and therefore Eq. \ref{partition} can be rewritten as:
\begin{equation}\label{partition2}
Z_{q}(\epsilon)\,\propto\,\epsilon^{(q-1)D_{q}}
\end{equation}
where $D_{q}$ is the spectrum of generalized dimensions. Note that the factor $q-1$ in the exponent ensures the validity of the normalization condition $Z_{1}(\epsilon)=1$.
Inside each $\epsilon\times\epsilon$ box, the local contribution of the multifractal measure $\mu$ is assumed to scale according to:
\begin{equation} \label{alphai}
\mu_{i}\,\propto\,\epsilon^{\alpha_{i}}
\end{equation}
where the local scaling exponent $\alpha_{i}=\alpha_{i}(\epsilon)$, also called the crowding index, is generally a position dependent quantity.
Furthermore, the number of boxes with a given crowding index $\alpha$ can be expressed as:
\begin{equation}\label{falpha}
N_{\alpha}(\epsilon)\,\propto\,\epsilon^{-f(\alpha)}
\end{equation}
Therefore, these boxes cover a subset with the fractal dimension $f(\alpha)$. At this point, we can immediately establish the following relation between the scaling function $\tau(q)$, the generalized dimension, and the partition function:
\begin{equation}\label{extra1}
\tau(q)=(1-q)D_{q}=-\lim_{\epsilon\rightarrow{0}}\frac{\log{Z_{q}(\epsilon)}}{\log{\epsilon}}
\end{equation} 
Moreover, by using Eq. \ref{alphai} and Eq. \ref{falpha} we can rewrite the partition function as:
\begin{equation}\label{partition3}
Z_{q}(\epsilon)\,\propto\,\int\epsilon^{\alpha{q}-f(\alpha)}d\alpha
\end{equation}
where the integral above is slowly varying over the smallest scales. In the limit $\epsilon\rightarrow{0}$, the integral above is dominated by the $\alpha$ values that minimize the exponent and it can be approximated using the saddlepoint method. Therefore, when $D_{q}$ and $f(\alpha)$ are differentiable functions we can  obtain:
\begin{equation}\label{impo1}
f(\alpha)=\alpha{q}-(q-1)D_{q}
\end{equation}
where $\alpha$ is given by:
\begin{equation}\label{impo2}
\alpha=\alpha(q)=\frac{d}{dq}[(q-1)D_{q}]
\end{equation}
The results above show again how the spectrum of generalized dimensions $D_{q}$ (or $\tau_{q}$) and the singularity (i.e., the multifractal) spectrum $f(\alpha)$ are connected by the Legendre transform and offer equivalent descriptions of the multifractal.

Finally, the generalized dimension $D_{q}$ can be related to the partition function as follows:
\begin{equation}\label{extra2}
D_{q}=\frac{1}{1-q}\lim_{\epsilon\rightarrow{0}}\frac{\log{Z_{q}(\epsilon)}}{-\log{\epsilon}}=\frac{1}{1-q}
\lim_{\epsilon\rightarrow{0}}\frac{\log{\sum_{i=1}^{N}\mu_{i}^{q}(\epsilon)}}{-\log{\epsilon}}
\end{equation}
where we recall that the local measure is the relative mass of the object in the $i$-th box, i.e., $\mu_{i}=M_{i}(\epsilon)/M$, where $M_{i}$ is the mass of the $i$-th box and $M$ is the total mass. For $q=0$ the expression above yields the box-counting dimension, for $q=1$ the information dimension, for $q=2$ the two-point correlation dimension. The generalized dimensions for $q>2$ provides information about higher-order correlations. For example, $D_{3}$ characterizes the correlations between triples of points in each box, $D_{4}$ between qaudruples, etc.

\section{Single scattering properties}\label{diffraction}
The fractal dimension $d_{f}$ of a structure can be obtained directly by measuring its correlation function, which is an observable quantity of fundamental importance in wave scattering experiments (e.g., in light, X-ray, and neutron scattering) \cite{anitas2019small}. The density-density correlation function is defined as:
\begin{equation}
g(\textbf{r},\textbf{r}^{\prime})=\langle\rho(\textbf{r})\rho(\textbf{r}^{\prime})\rangle
\end{equation}
where $\rho(\textbf{r})$ is the number density of atoms at position $\textbf{r}$ and $\langle\ldots\rangle$ denotes an ensemble average. The expression above quantifies the correlations in the fluctuation of the number density.
For isotropic atomic distributions, the correlation function depends on the radial distance $r=|\textbf{r}-\textbf{r}^{\prime}|$, which is often defined in spherical coordinates. Additionally, for systems that display translational invariance on average (i.e., statistically homogeneous), we can fix $\textbf{r}^{\prime}=0$ and write:
\begin{equation}
g(r)=\langle\rho(r)\rho(0)\rangle
\end{equation}
We note that the $g(r)$ defined above, known as the pair correlation function, is proportional to the probability of finding a particle at a distance $r$ from another particle of the system, which is proportional to the particle density $\rho(r)$ within a sphere of radius $r$.
Since $\rho(r)=dM(r)/dV\,\propto\,r^{d_{f}-d}$ for a monofractal distribution, it follows that the correlation function must scale as:
\begin{equation}
G(r)\,\propto\,r^{d_{f}-d}
\end{equation}
where $d$ is the dimension of the embedding Euclidean space.
The scattering intensity observed in actual experiments is proportional to the structure factor $S(q)$, which is essentially the Fourier transform of the pair correlation function:
\begin{equation}
S(q)=1+\frac{N}{V}\int_{V}[G(r)-1]e^{-i\textbf{q}\cdot\textbf{r}}d\textbf{r}
\end{equation}
where $N$ is the total number of particles in the system of volume $V$ and $q=\textbf{k}-\textbf{k}^{\prime}=(4\pi/\lambda)\sin(\theta/2)$ is the momentum transfer. Note that the momentum transfer should not be confused with the parameter $q$ used in the multifractal analysis. Here $\theta$ is the angle between the wavevectors $\textbf{k}$ and $\textbf{k}^{\prime}$.

The scaling of the structure factor based on a pair distribution function with a generic power-law singularity $G(r)\,\propto\,{r}^{-\alpha}$ can be easily estimated by noting that:
\begin{multline}
S(q)\propto\int{e^{-i\textbf{q}\cdot\textbf{r}}}d^{d}r/r^{\alpha}=(q^{\alpha}/q^{d})\int{e^{-i\textbf{q}\cdot\textbf{r}}}[d^{d}(qr)/(qr)^{\alpha}]\\\,\propto\,q^{\alpha-d}\int{e^{-ix}x^{-\alpha}d^{d}x}
\end{multline}
where $x=qr$. Remembering that in our case $-\alpha=d_{f}-d$, we obtain immediately:
\begin{equation}\label{structnice}
S(q)\,\propto\,q^{-d_{f}}
\end{equation}
More sophisticated models for the structure factor of fractals that include finite-size effects have been developed and are discussed in \cite{freltoft1986power,dimon1986structure,pearson1993long}. 

\begin{figure}[t!]
	\centering
	\includegraphics[width=\linewidth]{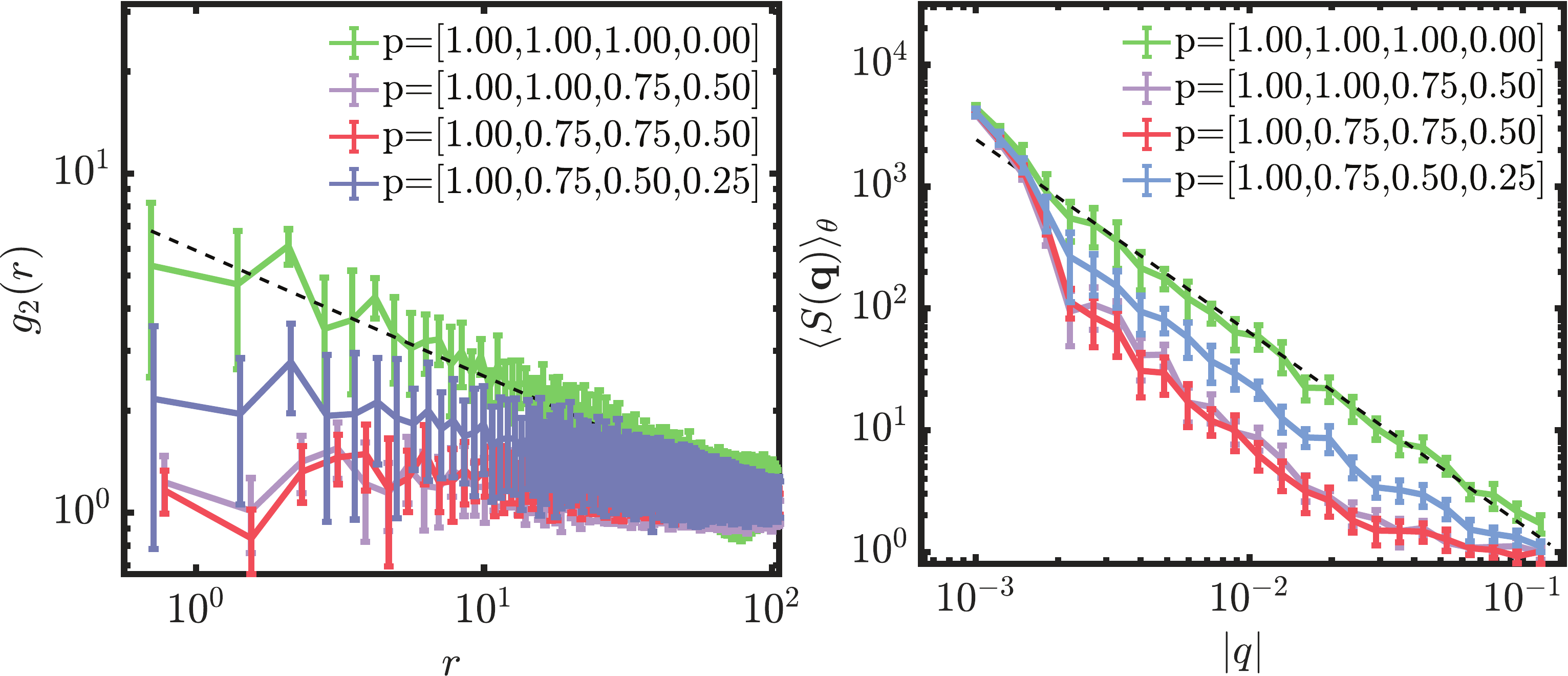}
    \caption{(a) Azimuthally averaged structure factor and (b) radial distribution function $g_2(r)$, which give the probability of finding two particles separated by a distance $r$, over 25 realizations for multifractal structures generated with the probability vectors specified in the legend.}
	\label{Fig4}
\end{figure}

\begin{figure*}[t!]
	\centering
	\includegraphics[width=\linewidth]{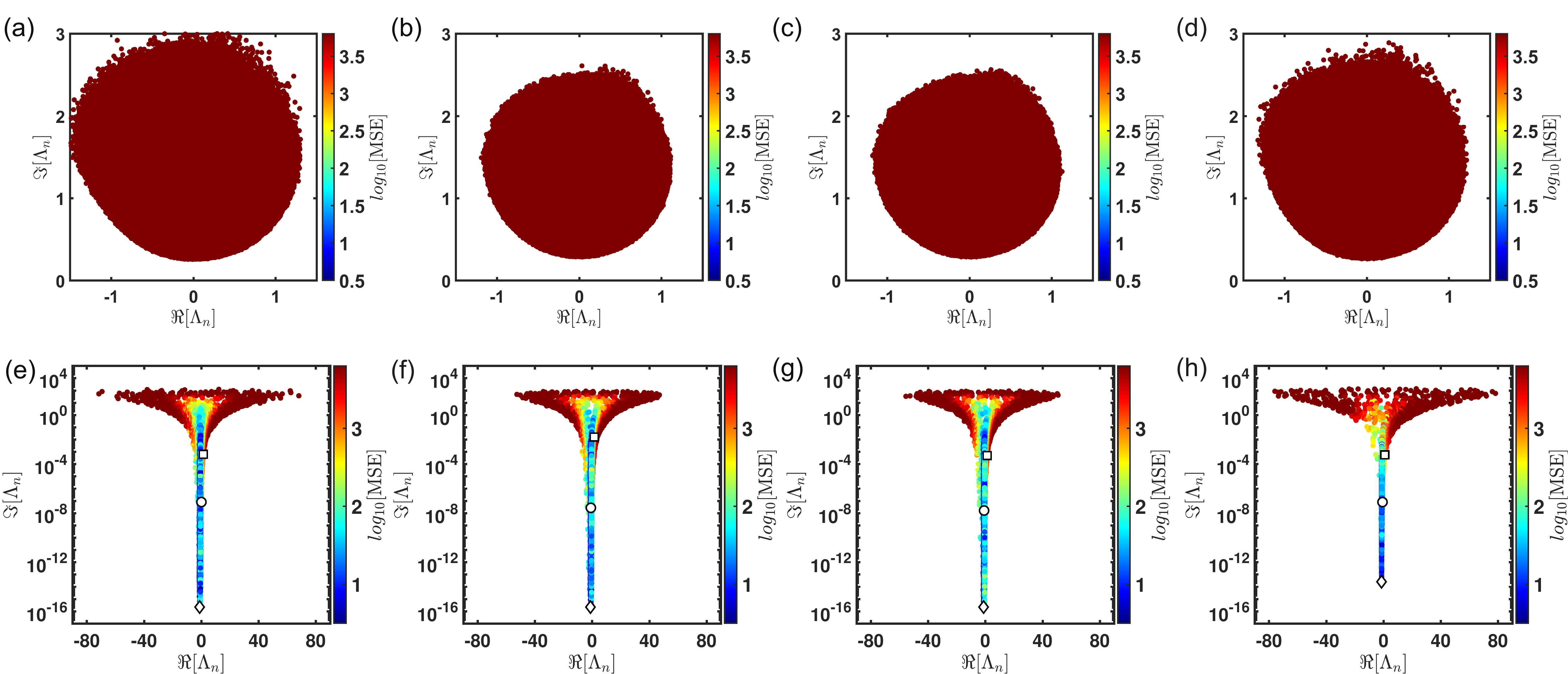}
    \caption{Panel (a-h) show the complex eigenvalue distribution, color coded with respect to the MSE parameter, of 25 different realizations of the point patterns presented in Fig.\ref{Fig1}. Specifically, panels (a-d) and panels (e-h) refer to two different optical regimes characterized by a value of $\rho\lambda^2$ equal to $10^{-6}$ and 50, respectively. The different markers in the panels (e-h) identify the spectral positions of the representative scattering resonances reported in Fig.\ref{Fig5}.}
	\label{Fig5}
\end{figure*}

\begin{figure*}[t!]
	\centering
	\includegraphics[width=\linewidth]{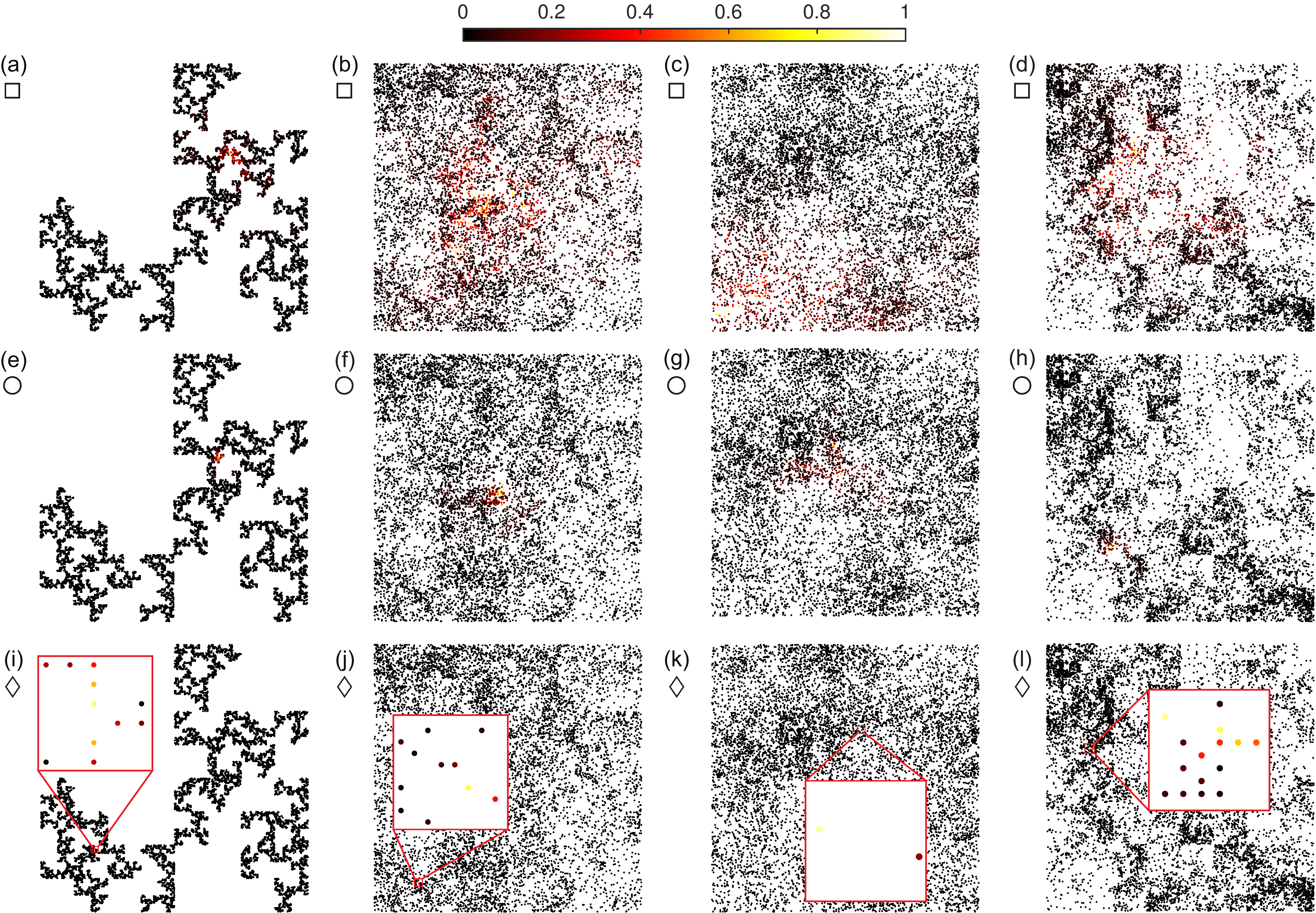}
    \caption{Representative scattering resonances of the investigated structures in the multiple scattering regime. While panels (a,e,i) refer to the monofractal point pattern generated with a probability $p=[1,1,1,0]$, panels (b,f,j), (c,g,k), and (d,h,l) display critical and Efimov-type few-body scattering resonances characterizing multifractal point patterns when $p$ is equal to  $[1,1,0.75,0.5]$, $[1,0.75,0.75,0.5]$, and $[1,0.75,0.5,0.25]$, respectively.}
\label{Fig6}
\end{figure*}

\begin{figure*}[t!]
	\centering
	\includegraphics[width=\linewidth]{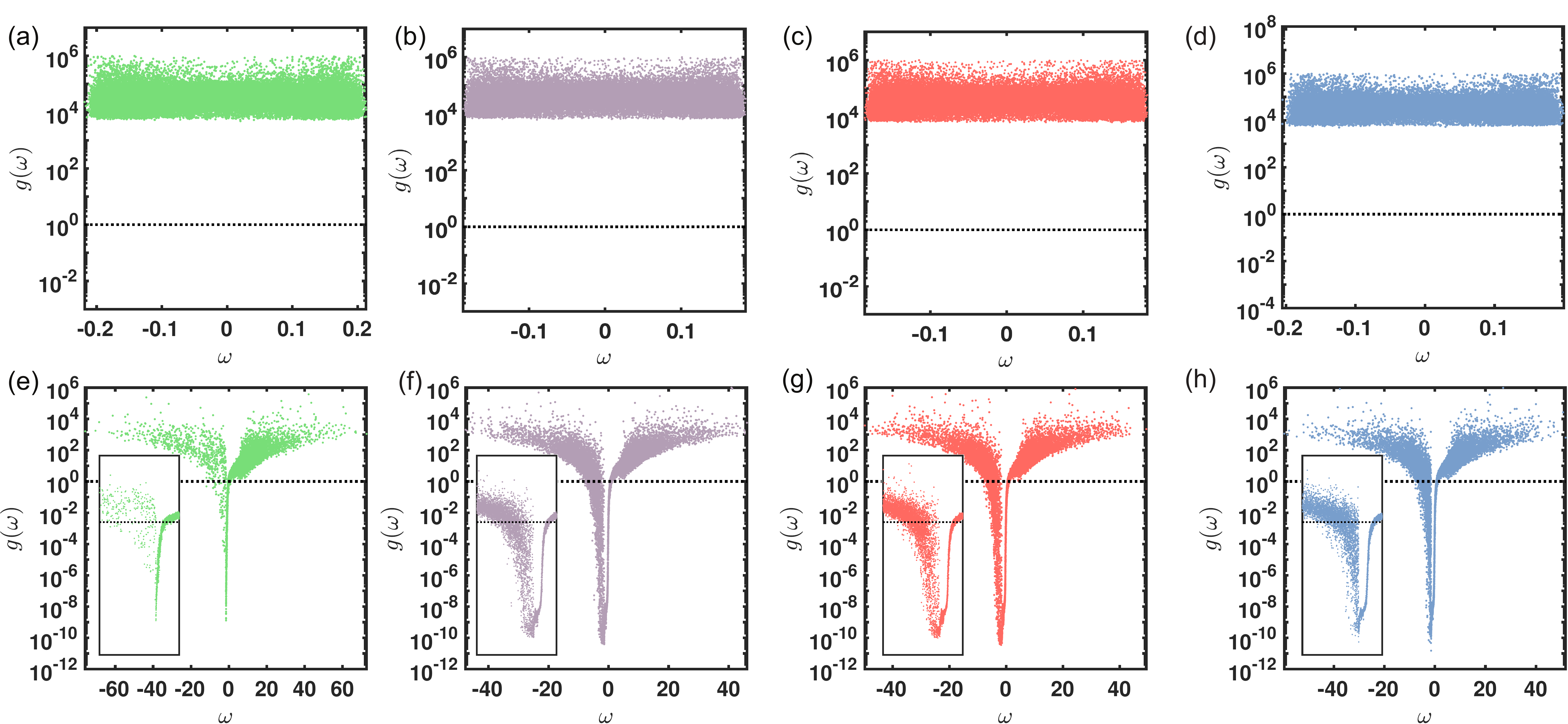}
    \caption{Panels (a-d) and (e-h) display the Thouless number as a function of the frequency $\omega$ extrapolated from the distribution reported in Fig.\ref{Fig4}\,(a-d)\,and\,(e-h), respectively. The pastel green, violet, red, and blue dots correspond to the Thouless number of 25 different disorder realizations produced when $\rho\lambda^2$ is equal to $10^{-6}$ (i.e, panels (a-d)) and to 50 (i.e, panels (e-h)), respectively. The dashed-black lines identify the threshold of the diffusion-localization transition. Panels (e-h) show a zoom-in view in the range $\omega\in[-10, 2]$.}
	\label{Fig7}
\end{figure*}

\begin{figure*}[t!]
	\centering
	\includegraphics[width=\linewidth]{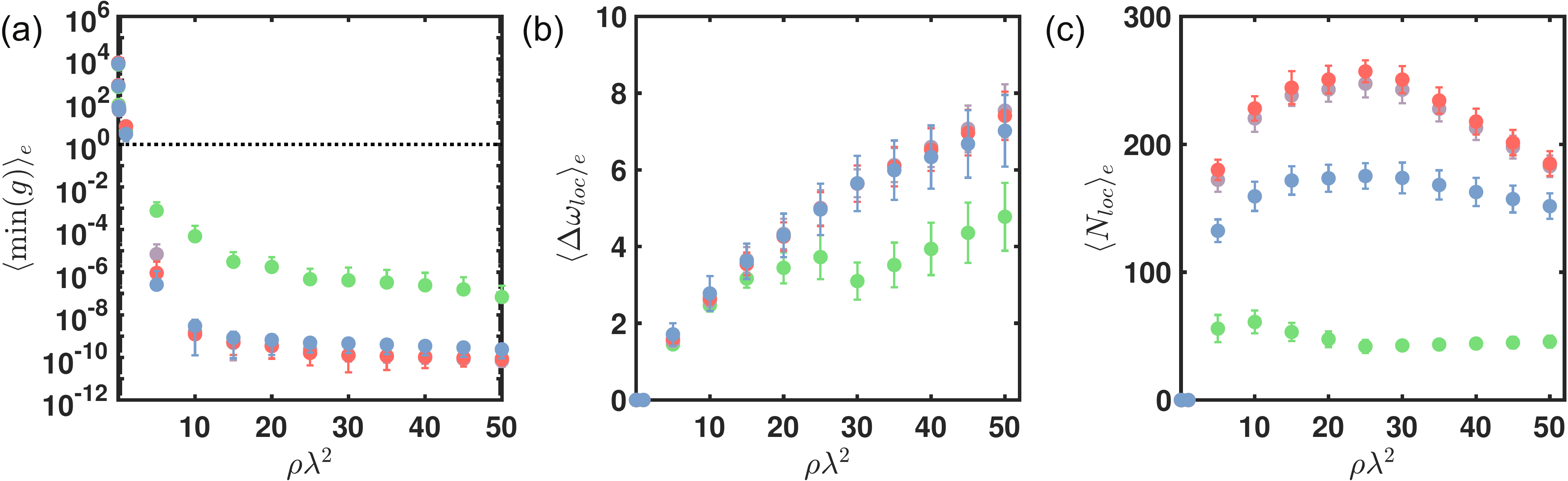}
    \caption{Panel (a) shows the trend of the minimum of the Thouless number as a function of $\rho\lambda^2$. Panel (b) displays the width of the  frequency range $\Delta\omega_{loc}$ for different optical densities where all the scattering resonances have a Thouless number lower than 1. Panel (c) shows the number of localized scattering resonances as a function of $\rho\lambda^2$.  All these parameters are averaged with respect to 25 different realizations and the error bars are the statistical errors associated with this average ensemble operation. The pastel green, violet, red, and blue markers refers to point patterns generated with a probability $p$ equal to $[1,1,1,0]$, $[1,1,0.75,0.5]$, $[1,0.75,0.75,0.5]$, and $[1,0.75,0.5,0.25]$, respectively.}
	\label{Fig8}
\end{figure*}

\begin{figure*}[t!]
	\centering
	\includegraphics[width=\linewidth]{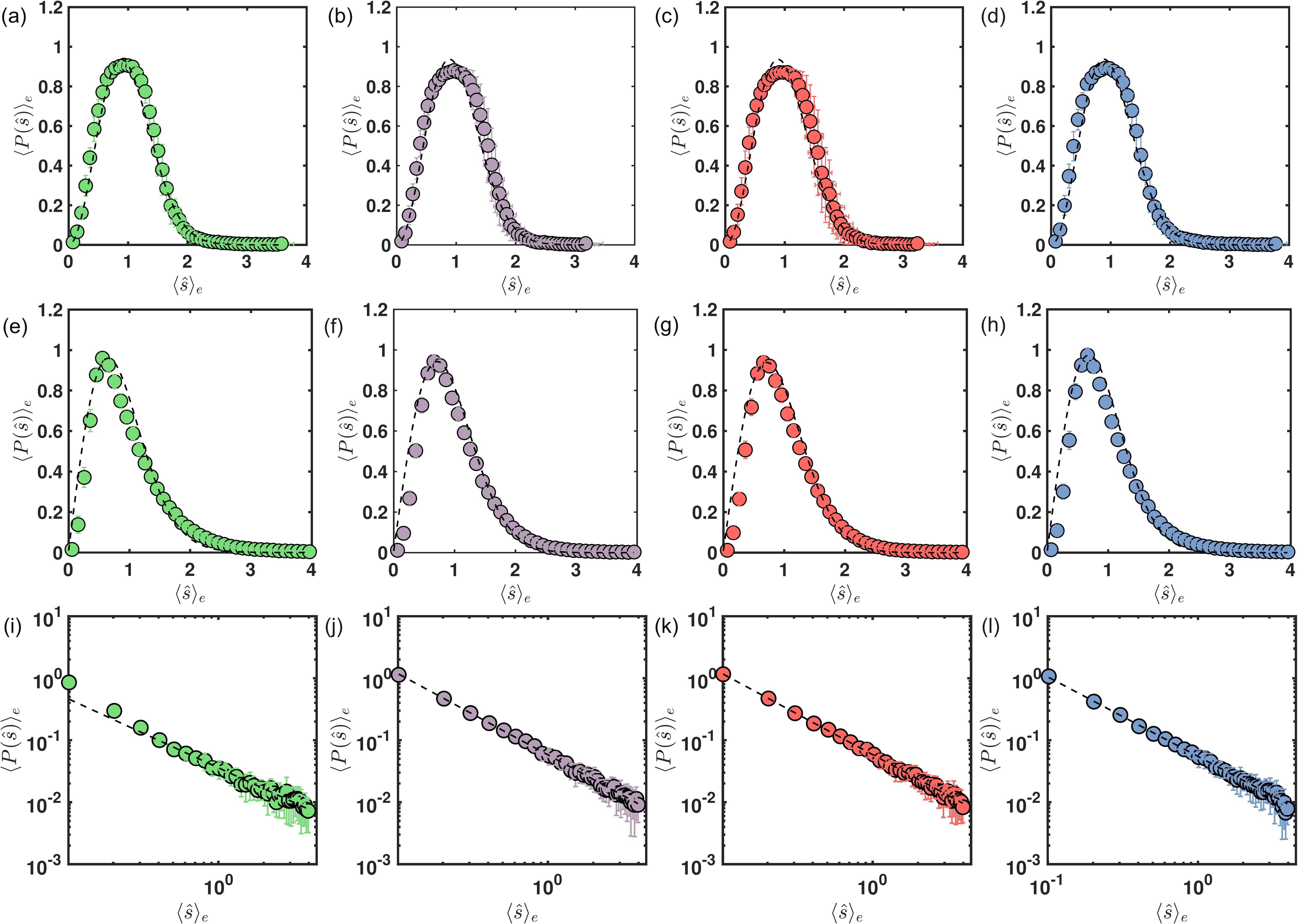}
    \caption{Ensemble averaged level spacing distribution $P(\hat{s})$ as a function of the nearest-neighbor Euclidean distance of the complex eigenvalues $|\Delta\Lambda|$=$|\Lambda_{n+1}-\Lambda_{n}|$ normalized to their average value, i.e, $\hat{s}$=$|\Delta\Lambda|/\langle|\Delta\Lambda|\rangle$. Panels (a-d), (e-h), and (i-l) correspond to a low (i.e, $\rho\lambda^2 = 10^{-6}$), intermediate (i.e, $\rho\lambda^2 = 10^{-1}$), and high (i.e, $\rho\lambda^2 = 50$) optical density regime, respectively. The black-dotted lines in panels (a-d), (e-h), and (i-l) identifies the different statistics that better describe the evolution of $\langle P(\hat{s})\rangle_e$ by increasing $\rho\lambda^2$ (see the main text for more details). The pastel green, violet, red, and blue markers refers to point patterns generated with a probability $p$ equal to $[1,1,1,0]$, $[1,1,0.75,0.5]$, $[1,0.75,0.75,0.5]$, and $[1,0.75,0.5,0.25]$, respectively.}
	\label{Fig9}
\end{figure*}

We now concisely address the role of structural correlations in the diffraction properties of multifractals. For simplicity we consider a one-dimensional measure $\mu(x)$ attached to some geometrical support of size $L$.
We would like to evaluate the two-point correlations of the moments of $\mu(x)$ defined by:
\begin{equation}\label{eq10}
G_{mn}(y)=\langle{\mu^{m}(x)\mu^{n}(x+y)}\rangle_{x}
\end{equation}
where the brackets indicate averaging over all sites $x$ and the correlation is between the local measures of boxes with fixed size $\epsilon$ and separation $y$. It has been shown in \cite{cates1987} that the correlation function of box measures can be described by exponents characterizing the multifractality of the set. In fact, due to the absence of characteristic length scales in the multifractal system, its pair correlation can be generally written as \cite{Nakayama}:
\begin{equation}
G_{mn}(y)\,\propto\,{\epsilon^{x_{1}(m,n)}L^{-x_{2}(m,n)y}y^{x_{3}(m,n)}}
\end{equation}
where the new exponents $x_{1},x_{2},x_{3}$ must be related to the previously introduced multifractal exponents. This is obtained by computing the correlation between boxes separated by the box-size distances $\epsilon$ and $L$, in which the boxes decorrelate. The steps of the derivation can be found in \cite{Nakayama} where the following expression for $x_{3}(m,n)$ has been obtained:
\begin{equation}
{x_{3}(m,n)=d_{f}-\tau(m)-\tau(n)+\tau(m+n)}
\end{equation}
where $d_{f}=D_{0}$ is the fractal dimension of the support (Note the opposite sign convention for $\tau(q)$ used here compared to the one in ref. \cite{Nakayama}). Therefore, the pair correlation  scales as \cite{cates1987}:
\begin{equation}\label{eq11}
G_{mn}(y)\,\propto\,{y^{d_{f}-\tau(m)-\tau(n)+\tau(m+n)}}
\end{equation}
Thus, the measurement of spatial correlations can provide an unambiguous quantitative test for multifractal behavior. 
Using the same argument that led to establish Eq. \ref{structnice}, we can deduce the general scaling of the structure factor of a multifractal for given $m$ and $n$ values:
\begin{equation}
S_{mn}(q)\,\propto\,q^{-d_{f}+\tau(m)+\tau(n)-\tau(n+m)}
\end{equation}
The relation above is quite important because it allows one to characterize the multifractality of a structure without box-counting procedures in numerical calculations, directly by studying the scaling of the structure factor for any given values of $m$ and $n$. However, when spatial or angular averages are performed, as in Fig. \ref{Fig4}, the above formula cannot be directly applied and a more complex approach, beyond the scope of our paper, must be developed to account for the spatial non-homogeneity of $g_{2}(r)$ and of the structure factor.

%%%%%%%%%%%%%%%%%%		  Acknowledgments              		   %%%%%%%%%%%%%%%%%%%%
%%%%%%%%%%%%%%%%%%%%%%%%%%%%%%%%%%%%%%%%%%%%%%%%%%%%%%%%%%

%\section*{Author contributions}
%F.S. performed numerical calculations, data analysis, and organized the results with the help of Y. C. S.G. performed the experiments together with F.S. W.A.B. fabricated the samples. L.D.N. conceived and \rev{led} the work. F.S. and L.D.N. wrote the manuscript.
%%%%%%%%%%%%%%%%%%%%%%%%%%%%%%%%%%%%%%%%%%%%%%%%%%%%%%%%%%
%%%%%%%%%%%%%%%%%%		        Bibilo              		   %%%%%%%%%%%%%%%%%%%%
%%%%%%%%%%%%%%%%%%%%%%%%%%%%%%%%%%%%%%%%%%%%%%%%%%%%%%%%%%
\bibliographystyle{apsrev4-1}
%

% ==================================  Fig1 - Structures generation ==========================================================
%%%%%%%%%%%%%%%%%%%%%%%%%%%%%%%%%%%%%%%%%%%%%%%%%%%%%%%%%%%%%%%%%%%%%%%%%%%%%%%%%%%%%%%%%%%%%%%%%%%%%%%%%%%%%%%%%%%%%%%%%%%%%%%%%%%%%%%%%%%%%%%%%%%%%%%%%%
% ==================================  Fig2 - MF geometrical properties  ==========================================================
%%%%%%%%%%%%%%%%%%%%%%%%%%%%%%%%%%%%%%%%%%%%%%%%%%%%%%%%%%%%%%%%%%%%%%%%%%%%%%%%%%%%%%%%%%%%%%%%%%%%%%%%%%%%%%%%%%%%%%%%%%%%%%%%%%%%%%%%%%%%%%%%%%%%%%%%%%
% ==================================  Fig3 - Distr First Neigh  ==========================================================
%%%%%%%%%%%%%%%%%%%%%%%%%%%%%%%%%%%%%%%%%%%%%%%%%%%%%%%%%%%%%%%%%%%%%%%%%%%%%%%%%%%%%%%%%%%%%%%%%%%%%%%%%%%%%%%%%%%%%%%%%%%%%%%%%%%%%%%%%%%%%%%%%%%%%%%%%%

% ==================================  Fig4 - g2 and Sk ==========================================================
%%%%%%%%%%%%%%%%%%%%%%%%%%%%%%%%%%%%%%%%%%%%%%%%%%%%%%%%%%%%%%%%%%%%%%%%%%%%%%%%%%%%%%%%%%%%%%%%%%%%%%%%%%%%%%%%%%%%%%%%%%%%%%%%%%%%%%%%%%%%%%%%%%%%%%%%%%

% ==================================  Fig5  -  Eig Distribution ==========================================================
%%%%%%%%%%%%%%%%%%%%%%%%%%%%%%%%%%%%%%%%%%%%%%%%%%%%%%%%%%%%%%%%%%%%%%%%%%%%%%%%%%%%%%%%%%%%%%%%%%%%%%%%%%%%%%%%%%%%%%%%%%%%%%%%%%%%%%%%%%%%%%%%%%%%%%%%%%

% ==================================  Fig5 - Representative eigenvectors  ==========================================================
%%%%%%%%%%%%%%%%%%%%%%%%%%%%%%%%%%%%%%%%%%%%%%%%%%%%%%%%%%%%%%%%%%%%%%%%%%%%%%%%%%%%%%%%%%%%%%%%%%%%%%%%%%%%%%%%%%%%%%%%%%%%%%%%%%%%%%%%%%%%%%%%%%%%%%%%%%

% ==================================  Fig6 - Thouless number  ==========================================================
%%%%%%%%%%%%%%%%%%%%%%%%%%%%%%%%%%%%%%%%%%%%%%%%%%%%%%%%%%%%%%%%%%%%%%%%%%%%%%%%%%%%%%%%%%%%%%%%%%%%%%%%%%%%%%%%%%%%%%%%%%%%%%%%%%%%%%%%%%%%%%%%%%%%%%%%%%

% ==================================  Fig7 - Min Thouless & ect ==========================================================
%%%%%%%%%%%%%%%%%%%%%%%%%%%%%%%%%%%%%%%%%%%%%%%%%%%%%%%%%%%%%%%%%%%%%%%%%%%%%%%%%%%%%%%%%%%%%%%%%%%%%%%%%%%%%%%%%%%%%%%%%%%%%%%%%%%%%%%%%%%%%%%%%%%%%%%%%%
% ==================================  Fig8 -Level spacing statistics  ==========================================================
%%%%%%%%%%%%%%%%%%%%%%%%%%%%%%%%%%%%%%%%%%%%%%%%%%%%%%%%%%%%%%%%%%%%%%%%%%%%%%%%%%%%%%%%%%%%%%%%%%%%%%%%%%%%%%%%%%%%%%%%%%%%%%%%%%%%%%%%%%%%%%%%%%%%%%%%%%

\end{document}